\documentclass[%
 reprint,
 amsmath,amssymb,
 aps,
 ]{revtex4-2}

\usepackage{graphicx}% Include figure files
\usepackage{dcolumn}% Align table columns on decimal point
\usepackage{bm}% bold math
\usepackage[pdfstartview=FitH,
CJKbookmarks=true,
bookmarksnumbered=true,
bookmarksopen=true,
colorlinks,
linkcolor=blue,
anchorcolor=blue,
citecolor=blue,
allcolors=blue,
pdfborder=001,
]{hyperref}
\usepackage[caption=false]{subfig}
\usepackage{floatrow}
\usepackage{enumitem}
\DeclareSubrefFormat{myparens}{#1~(#2)}
\DeclareCaptionListOfFormat{myparens}{#1(#2)}
\usepackage{xspace}
\usepackage{multirow}
\usepackage{makecell}
\usepackage{placeins}

\begin{document}

\preprint{APS/123-QED}

\newcommand{\BESIIIorcid}[1]{\href{https://orcid.org/#1}{\hspace*{0.1em}\raisebox{-0.45ex}{\includegraphics[width=1em]{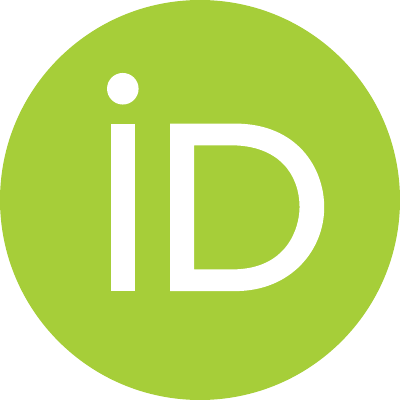}}}} 

\newcommand\MeV{\ensuremath{\mathrm{MeV}}}
\newcommand\GeV{\ensuremath{\mathrm{GeV}}}

\setlength{\abovedisplayskip}{6pt}
\setlength{\belowdisplayskip}{6pt}

\title{\boldmath Lightest $0^{-+}$ Glueball as Dominant Constituent of $X(2370)$}

\author{BESIII Collaboration}
\altaffiliation{Full author list given at the end of the Letter.}

{
\renewcommand{\baselinestretch}{1.20}

\begin{abstract} 
With 10 billion $J/\psi$ events collected at \mbox{BESIII}, the decay
of $X(2370)\rightarrow K^{*}(892)^{0}\bar{K}^{0}+\mathrm{c.c.}$ is
searched for via the $J/\psi\rightarrow\gamma K_{S}^{0}K_{S}^{0}\pi^{0}$ process. 
No evidence of this decay mode is found, with $\mathcal{B}[J/\psi\rightarrow\gamma X(2370)] \times \mathcal{B}[ X(2370) \rightarrow K^{*}(892)^{0}\bar{K}^{0}+\mathrm{c.c.} \rightarrow K^{0}_{S}K^{0}_{S}\pi^{0}] < 2.7\times 10^{-6}$ at the $90\%$ confidence level. 
The suppression of the $K^{*}(892)\bar{K}$ mode indicates the $X(2370)$ as a flavor-singlet state. 
It is the first flavor-singlet light hadron observed above $1~\GeV/c^{2}$. 
All \mbox{BESIII} measurements on the $X(2370)$ are summarized. 
A dominant component of the lightest $0^{-+}$ glueball is essential for a natural and complete explanation of the properties of the $X(2370)$~\textemdash~the mass, spin-parity, high production rate in $J/\psi$ radiative decays, decay pattern similarities to that of $\eta_{c}$, flavor-singlet property, narrow partial decay width, and suppression of radiative decays to $\omega$ and $\phi$, since these properties are all consistent with the features of the lightest $0^{-+}$ glueball while other interpretations at present are disfavored.
This supports that the lightest $0^{-+}$ glueball is the dominant constituent of the $X(2370)$.

\end{abstract}

\maketitle

In the Standard Model, the strong interaction is mediated by gluons and is described by the non-Abelian SU(3) gauge theory, 
which is known as the Quantum Chromodynamics (QCD). The non-Abelian gauge property predicts self-interactions among gluons, 
which allow gluons to form bound states~\textemdash~glueballs. 
As a unique type of matter made of gauge bosons~\textemdash~force carriers, 
the existence of glueballs is an important prediction of QCD at low energy. 
The search for glueballs is a direct test of the non-Abelian nature
of the strong interaction. 
For the $0^{-+}$ ground state glueball, 
lattice QCD (LQCD) predicts its mass to be between $2.3$ and $3.0~\GeV/c^{2}$~\cite{LQCD1,LQCD2,LQCD3,LQCD4,LQCD5,Vadacchino:2023vnc}.

Glueballs are expected to be copiously produced in gluon-rich environments, 
and $J/\psi$ radiative decays are believed to be an ideal place to search for glueballs~\cite{Kopke:1988cs,PhysRevD.50.3268,PhysRevD.55.5749}. 
The LQCD calculation predicts that the production rate of $J/\psi$ radiative decays to $0^{-+}$ glueball is $(2.31\pm0.90)\times10^{-4}$~\cite{LQCD5}.
A mixing mechanism between the $0^{-+}$ glueball and
$c\bar{c}$, which is also predicted by LQCD ~\cite{Zhang:2021xvl}, can significantly increase the
production rate of the mixture particle in $J/\psi$ radiative decays
to $(2\text{--}20)\times10^{-3}$, even with a tiny mixing angle
$2^{\circ}$--$5^{\circ}$ between the glueball and the $c\bar{c}$ component~\cite{Chen:2026lki}. 
Such a mixing mechanism means that $0^{-+}$ glueball might only be
found in the $0^{-+}$ glueball and $c\bar{c}$ mixture, i.e.,
identification of the $0^{-+}$ glueball dominant mixture particle should
be an unambiguous proof of the existence of the $0^{-+}$ glueball. The
decay properties of this mixture particle will keep almost all those
of the $0^{-+}$ glueball due to the tininess of the mixing and similar
decay patterns of the glueball to those of a $c\bar{c}$ charmonium with the same $J^{PC}$, since both of them predominantly decay via gluons~\cite{Chao:1995hd,Huang:1995td,Huang:2025pyv,Morningstar:2024vjk}.
In particular, according to the Okubo--Zweig--Iizuka (OZI) rule, both $0^{-+}$ glueball decays and $\eta_{c}$ decays are suppressed~\cite{Robson:1977pm}, 
therefore, they have no dominant decay modes and the partial decay widths of each decay mode should be small. 
Glueballs are flavor-singlet states, i.e., they contain no quark flavor content, so they are expected to decay flavor symmetrically into light hadron final states. 
The radiative decays of mesons to $\omega$ and $\phi$ are believed as important tags of $u\bar{u}+d\bar{d}$ and $s\bar{s}$ contents inside them~\cite{CHANOWITZ1985379,Close:2002sf,Hechenberger:2023ljn}. Thus, strong suppressions of such decays are expected for glueballs, since they contain no quarks,
and have no direct interaction with photons. 
Although no single criterion could unambiguously distinguish the glueballs from other kinds of hadrons, 
and glueballs may also have other properties, all above production and decay properties are the most important signatures of
glueballs.
Synthesized experimental studies on all these properties are necessary and will enable to identify whether a particle is a glueball or not because other kinds of hadrons can hardly satisfy all of these criteria.

The $X(2370)$ particle has been regarded as a good candidate for the lightest $0^{-+}$ glueball since it was first observed in the process $J/\psi\rightarrow\gamma \pi^{+}\pi^{-}\eta^{\prime}$ at \mbox{BESIII} experiment in 2011~\cite{1835_confirmed}. In this Letter, we summarize comprehensive studies on the $X(2370)$ performed by the \mbox{BESIII}, including the first determination on the $X(2370)$ flavor-singlet property, and examine whether all measured properties of the $X(2370)$ are consistent with the above features of glueballs or with other interpretations~\cite{Sun:2021kka,Yu:2011ta,Dong:2020okt,Wang:2025nme,Su:2022eun}.

With 225 million $J/\psi$ events collected at \mbox{BESIII}, the $X(2370)$ was observed in the process $J/\psi\rightarrow\gamma \pi^{+}\pi^{-}\eta^{\prime}$ with a statistical significance greater than $6.4\sigma$. Its mass and width were measured to be $2376.3 \pm 8.7(\mathrm{stat})^{+3.2}_{-4.3}(\mathrm{syst})~\MeV/c^{2}$ and $83\pm17(\mathrm{stat})^{+44}_{-6}(\mathrm{syst})~\MeV$~\cite{1835_confirmed}, respectively.
Its mass is consistent with the LQCD prediction for the lightest pseudoscalar glueball~\cite{LQCD1,LQCD2,LQCD3,LQCD4,LQCD5,Vadacchino:2023vnc}. 
The $X(2370)$ was confirmed in the decays $J/\psi\rightarrow \gamma K^{+}K^{-}\eta^{\prime}$ and $J/\psi\rightarrow \gamma K^{0}_{S}K^{0}_{S}\eta^{\prime}$ with a statistical significance of $8.3\sigma$~\cite{epjc.s10052_gongli}. 
Based on 10 billion $J/\psi$ events, 
the spin-parity quantum numbers of the $X(2370)$ were determined to be $0^{-+}$ for the first time via the $J/\psi\rightarrow\gamma K^{0}_{S}K^{0}_{S}\eta^{\prime}$ process with a statistical significance greater than $9.8\sigma$
~\cite{2370jpc}.
Recently, new decay modes of $X(2370)\rightarrow K_{S}^{0}K_{S}^{0}\pi^{0}$, $X(2370)\rightarrow\pi^{0}\pi^{0}\eta$ and $a_0(980)^{0}\pi^{0}$ were observed in the processes $J/\psi\rightarrow\gamma K_{S}^{0}K_{S}^{0}\pi^{0}$ and $J/\psi\rightarrow\gamma \pi^{0}\pi^{0}\eta$~\cite{combined_X2370}. The observed decay modes of the $X(2370)$ show decay pattern similarities to that of $\eta_c$. 
In addition, the $X(2370)$ decays into $\gamma\omega$ or $\gamma\phi$ are strongly suppressed in the processes $J/\psi\rightarrow\gamma\gamma\omega$ and  $J/\psi\rightarrow\gamma\gamma\phi$~\cite{BESIII:ggomega,BESIII:2024ein}.
All the above BESIII measurements are consistent with the expected properties of the lightest $0^{-+}$ glueball.

Flavor-singlet property is one of the most important properties of glueballs. 
The flavor symmetric decays of hadrons are usually examined by comparing branching fractions~(BFs) of different decay processes. However, for such comparisons, definite conclusions can hardly be drawn due to large uncertainties,
e.g., $\mathcal{B}[J/\psi\rightarrow\gamma X(2370)]\times\mathcal{B}[X(2370)\rightarrow f_{0}(980)\eta^{\prime}]\times\mathcal{B}[f_{0}(980)\rightarrow K^{0}_{S} K^{0}_{S}]=( 1.31 \pm 0.22 ({\rm stat})^{+2.85}_{-0.84}({\rm syst}) ) \times 10^{-5}$~\cite{2370jpc}.
Fortunately, the $K^{*}(892)^{0}\bar{K}^{0}$ mode suppression is an
unambiguous signature for a $0^{-+}$ flavor-singlet, since a $0^{-+}$
flavor-singlet meson is forbidden to decay into
$K^{*}(892)^{0}\bar{K}^{0}$ mode due to the generalized $G$-parity
conservation~\cite{LIPKIN1981114,LIPKIN1982326,KLEMPT20071}. Therefore,
searching for the $X(2370)\rightarrow K^{*}(892)^{0}\bar{K}^{0}$ would
be a crucial test whether it is a flavor-singlet, as expected for a glueball.

The decay of
$X(2370)\rightarrow K^{*}(892)^{0}\bar{K}^{0}+\mathrm{c.c.}$ is
searched for via the
$J/\psi\rightarrow\gamma K_{S}^{0}K_{S}^{0}\pi^{0}$ process, based on $(10087\pm44)\times10^{6}$ $J/\psi$
events~\cite{number} collected by the \mbox{BESIII}. A
detailed description of the \mbox{BESIII} detector can be found in
Ref.~\cite{detector}. Candidates for the process
$J/\psi\rightarrow\gamma K_{S}^{0}K_{S}^{0}\pi^{0}$ are selected using
the criteria described in Ref.~\cite{combined_X2370} and the
corresponding background contributions are estimated to be negligible.
After applying the selection criteria, a clear signature of the
$X(2370)$ around $2.3~\GeV/c^{2}$, along with the $\eta_{c}$ peak, is
observed in the $K^{0}_{S}K^{0}_{S}\pi^{0}$ invariant mass spectrum,
as shown in Fig.~\ref{fig:mkskspi0} as well as a clear $K^{*}(892)^{0}$ peak
is observed in the $K^{0}_{S}\pi^{0}$ invariant mass spectrum, shown in Fig.~\ref{fig:mpi0ks}.  The two-dimensional (2D) distribution
of $M_{K^{0}_{S}\pi^{0}}$ versus $M_{K^{0}_{S}K^{0}_{S}\pi^{0}}$ is
shown in Fig.~\ref{fig:2Dplot_pi0ks_kskspi0}.  To select the
$K^{*}(892)^{0}$ signal, each event is further required to have at
least one $K^{0}_{S}\pi^{0}$ combination in the $K^{*}(892)^{0}$ mass
region ($|M_{K^{0}_{S}\pi^{0}}-m_{K^{*}(892)^{0}}|\leq50~\MeV/c^{2}$,
where $m_{K^{*}(892)^{0}}$ is the nominal mass of the
$K^{*}(892)^{0}$~\cite{pdg2024}).  In the resulting
$K_{S}^{0}K_{S}^{0}\pi^{0}$ invariant mass spectrum, there are clear
mass peaks around $1.4~\GeV/c^{2}$ and $1.9~\GeV/c^{2}$, but no
evident $X(2370)$ peak, as shown in Fig.~\ref{fig:mkskspi0_cutkstar}.

\begin{figure}[tb]
\centering
\vspace{-5mm}
\subfloat{
\hspace{-2mm}
  \includegraphics[width=0.5\textwidth]{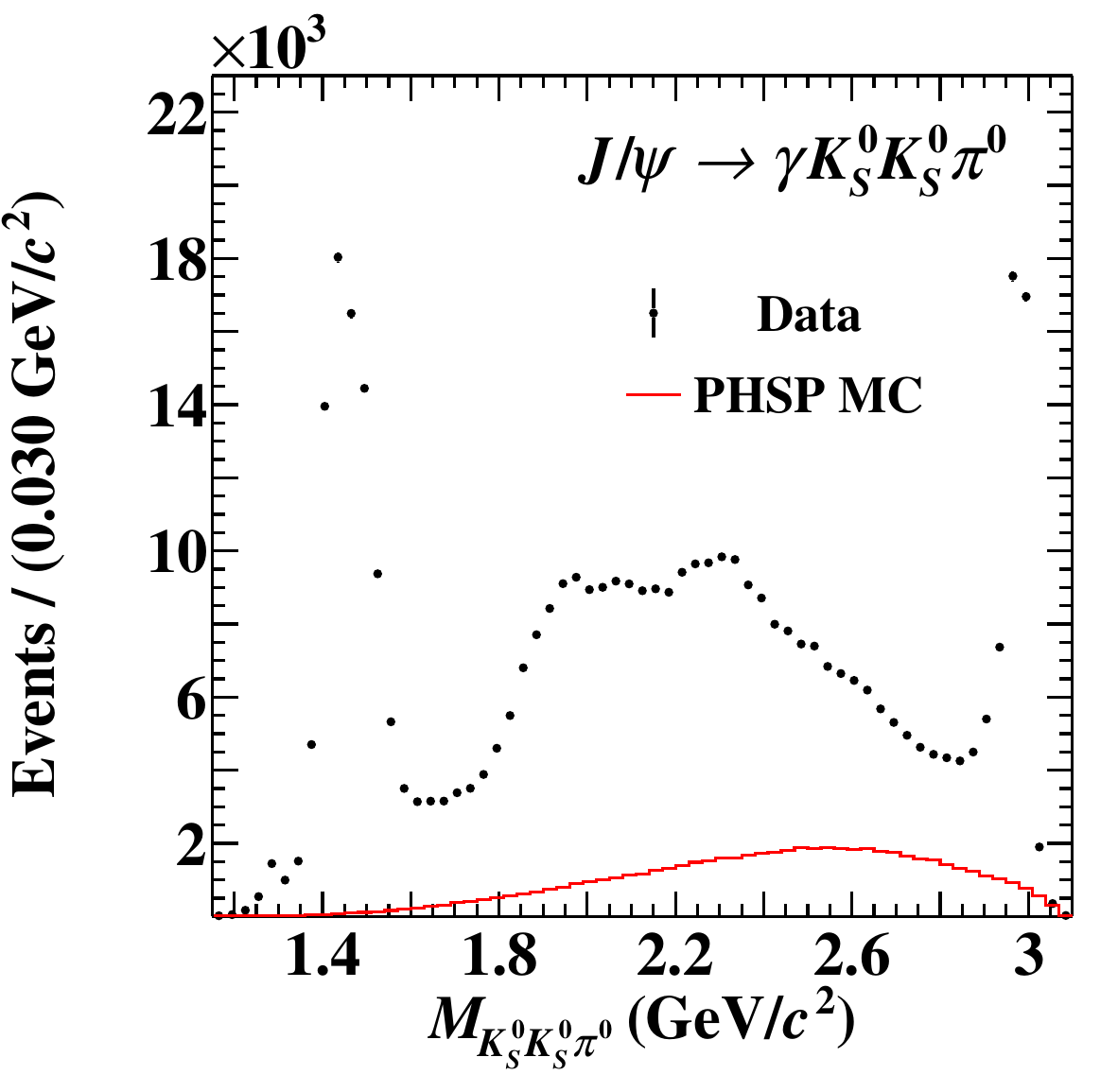}
  \label{fig:mkskspi0}
\put(-100,100){(a)}
}\subfloat{
\hspace{-2mm}
  \includegraphics[width=0.5\textwidth]{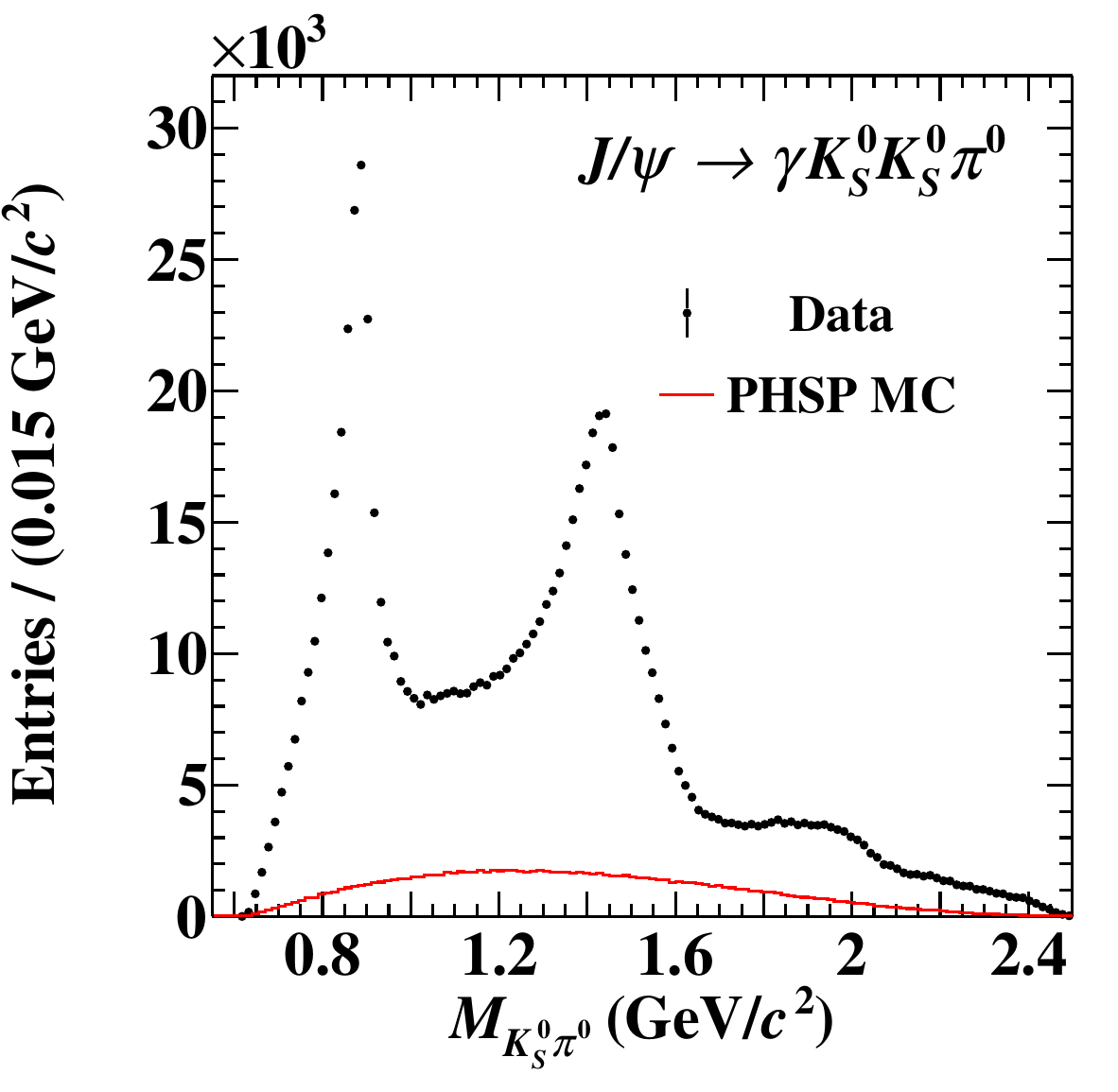}
  \label{fig:mpi0ks}
\put(-100,100){(b)}
}\\
\vspace{-4.5mm}
\subfloat{
\hspace{-2mm}
  \includegraphics[width=0.5\textwidth]{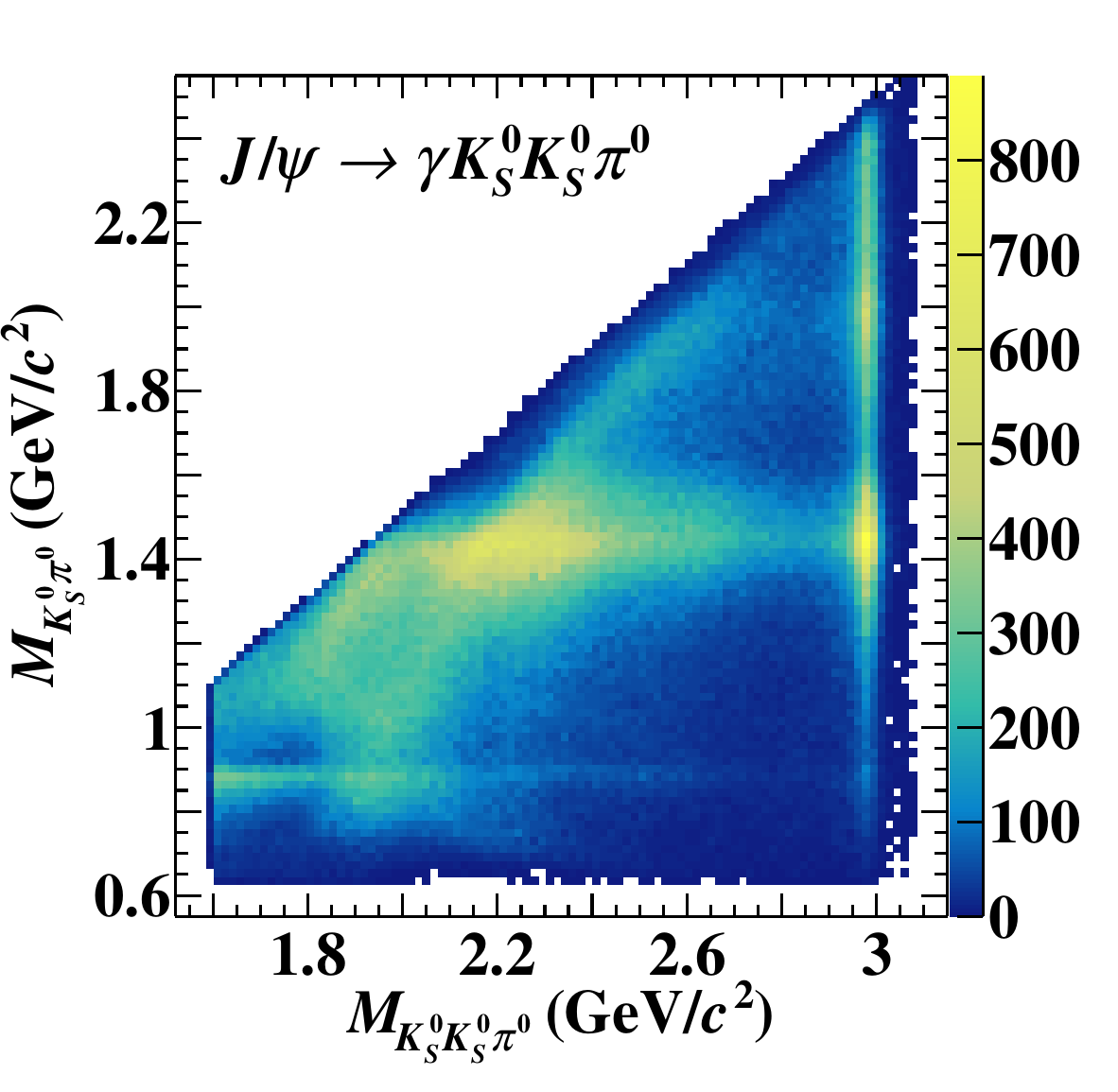}
  \label{fig:2Dplot_pi0ks_kskspi0}
\put(-100,86){(c)}
}
\subfloat{
\hspace{-2mm}
  \includegraphics[width=0.5\textwidth]{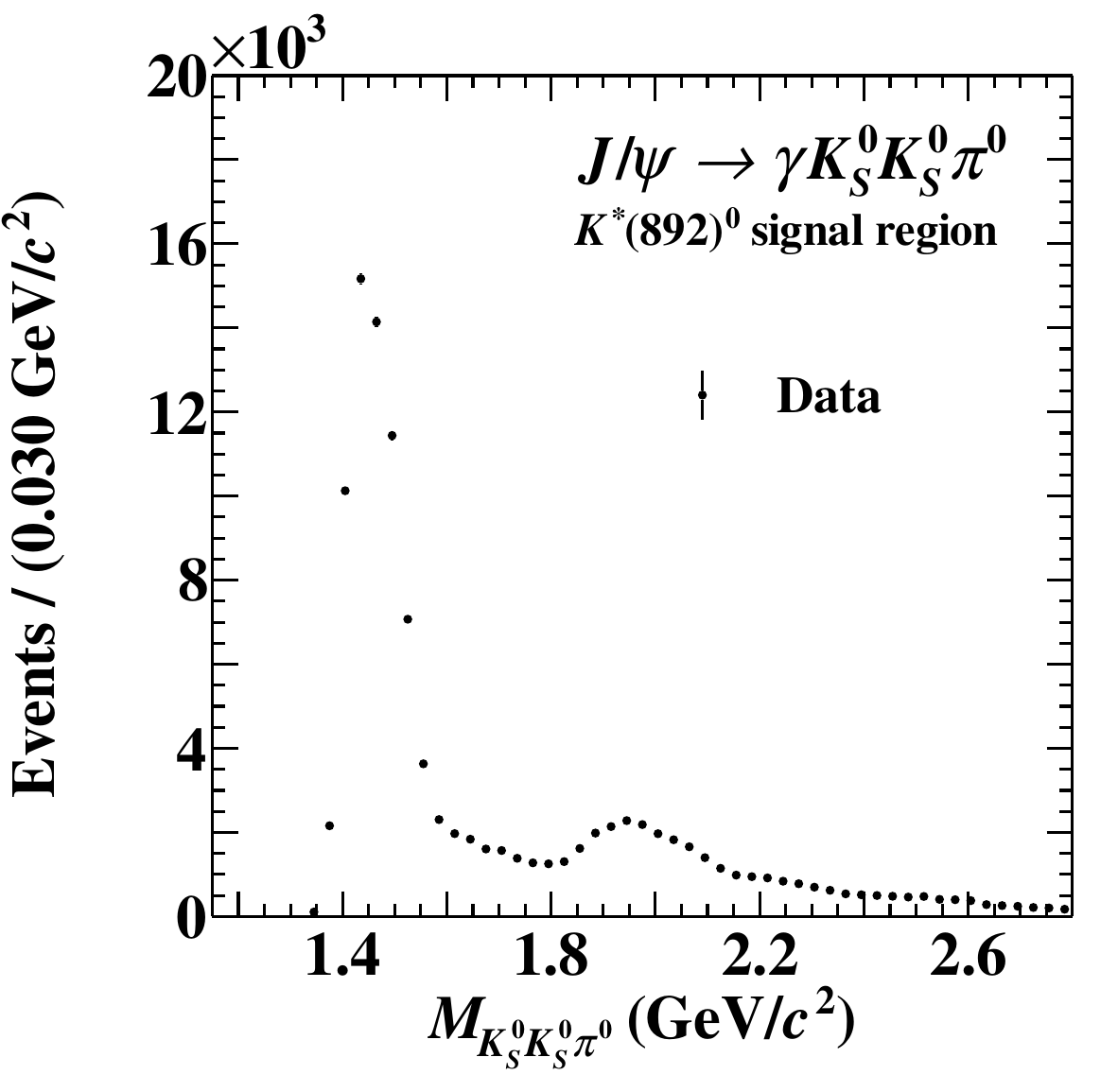}
  \label{fig:mkskspi0_cutkstar} 
\put(-100,100){(d)}
}\\
\vspace{-3.0mm}
\caption{
The (a) $K^{0}_{S}K^{0}_{S}\pi^{0}$ and (b) $K^{0}_{S}\pi^{0}$ (two
entries per event) invariant mass spectra. The points with error bars
are data and the solid red histograms are the PHSP simulated samples with arbitrary normalization.
(c) The 2D distribution of $M_{K^{0}_{S}\pi^{0}}$ (two entries per event) versus $M_{K^{0}_{S}K^{0}_{S}\pi^{0}}$.
(d) The $K^{0}_{S}K^{0}_{S}\pi^{0}$ invariant mass spectrum for selected events in the $K^{*}(892)^{0}$ mass region.}
  \label{fig:spectra_and_fit1}
\end{figure}

An unbinned maximum-likelihood fit is performed to the
$K^{0}_{S}K^{0}_{S}\pi^{0}$ invariant mass spectrum after the
$K^{*}(892)^{0}$ mass requirement to obtain the product BF $\mathcal{B}[J/\psi\rightarrow\gamma X(2370)] \times \mathcal{B}[ X(2370) \rightarrow K^{*}(892)^{0}\bar{K}^{0}+\mathrm{c.c.} \rightarrow K^{0}_{S}K^{0}_{S}\pi^{0}]$ (denoted as $\mathcal{B}_{K^{*}\bar{K}}$). 
The $X(2370)$ signal is described by an efficiency-corrected and phase space (PHSP) weighted
Breit--Wigner function convolved with a Gassian resolution function, where the efficiency and resolution are determined from Monte Carlo (MC) simulations.
A third-order polynomial is used to describe the contributions from the remaining processes. 
The mass and width of the $X(2370)$ are fixed to the combined results of $2359~\MeV/c^{2}$ and $170~\MeV$, respectively~\cite{combined_X2370}. 
The fit is shown in Fig.~\ref{fig:fit_mkskspi0_cutkstar}, and the statistical significance of the $X(2370)$ is $0.1\sigma$, and $\mathcal{B}_{K^{*}\bar{K}}$ is determined to be $(0.1\pm1.2(\mathrm{stat}))\times10^{-6}$. 
To obtain the BF ratio $\mathcal{R}=\frac{\mathcal{B}[X(2370) \rightarrow K^{*}(892)^{0}\bar{K}^{0}+\mathrm{c.c.}\rightarrow K_{S}^{0}K_{S}^{0}\pi^{0}]}{\mathcal{B}[ X(2370) \rightarrow K_{S}^{0}K_{S}^{0}\pi^{0}] }$, an unbinned maximum-likelihood fit is performed to the $K^{0}_{S}K^{0}_{S}\pi^{0}$ invariant mass spectrum without applying the $K^{*}(892)^{0}$ requirement.  
The description for the $X(2370)$ is analogous to that for obtaining $\mathcal{B}_{K^{*}\bar{K}}$, and its resonance parameters are fixed to the combined values~\cite{combined_X2370}, 
with the efficiency curve and PHSP being changed accordingly.  
Considering only the statistical uncertainty of $\mathcal{B}_{K^{*}\bar{K}}$, $\mathcal{R}$ is determined to be $0.003\pm0.040(\mathrm{stat})$.

\begin{figure}[tb]
\centering
\vspace{-5mm}
\hspace{-2mm}
  \includegraphics[width=0.65\textwidth]{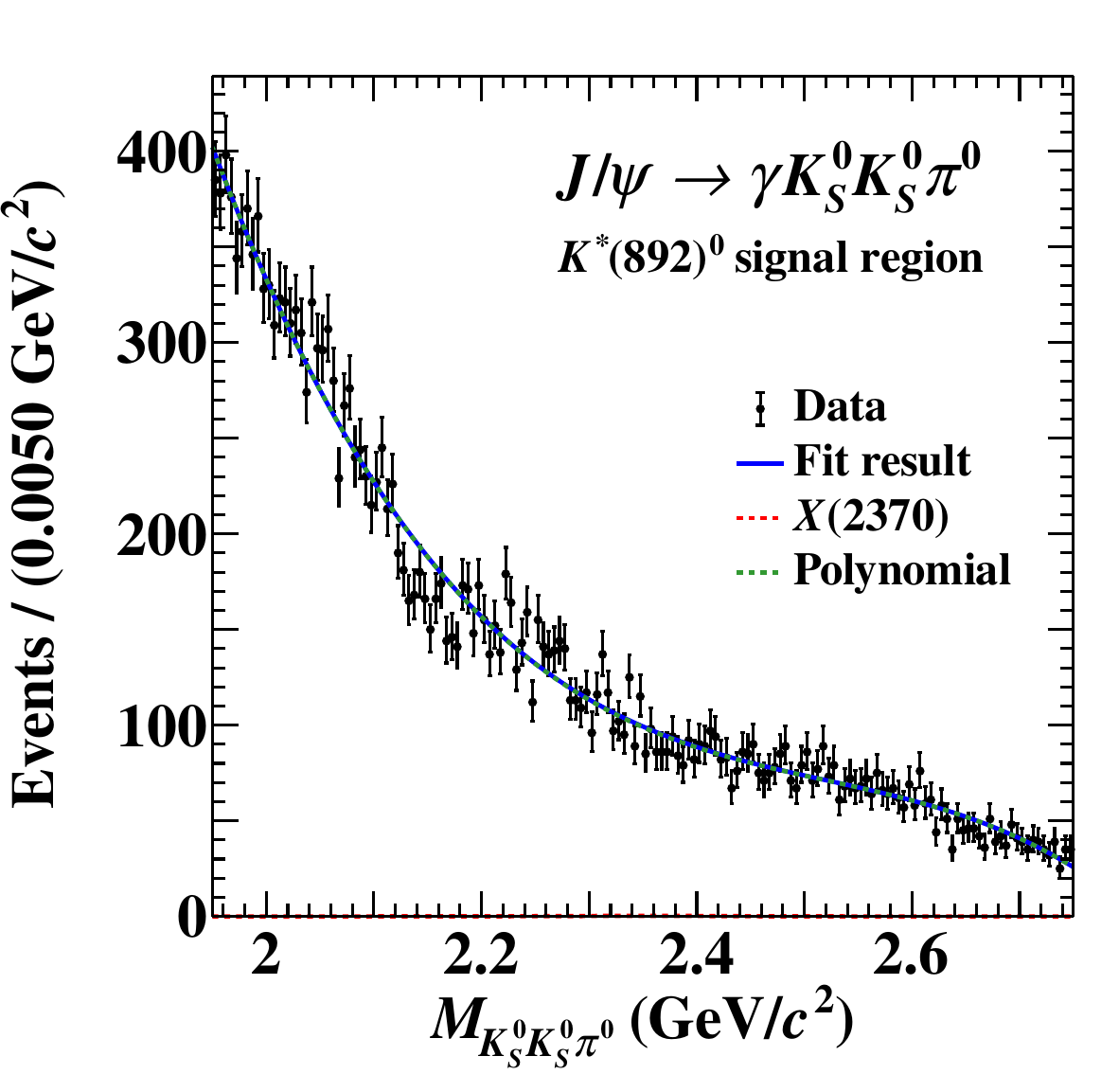}
  \label{fig:plot_mkspi0_mkskspi0}
\vspace{-3.0mm}
\caption{The fit result of the $K^{0}_{S}K^{0}_{S}\pi^{0}$ invariant mass spectrum for selected events in the $K^{*}(892)^{0}$ signal region. The dashed red line is the $X(2370)$ signal component, the dashed green line is the remaining processes described by a third-order polynomial function, and the blue line is the total fit result.
}
  \label{fig:fit_mkskspi0_cutkstar}
\end{figure}

The systematic uncertainties on $\mathcal{B}_{K^{*}\bar{K}}$ and $\mathcal{R}$ are evaluated by varying the $X(2370)$ parametrization, the PHSP parametrization, the background model, and possible additional resonances. The uncertainty from the event selection efficiency is additionally included for $\mathcal{B}_{K^{*}\bar{K}}$. 
With systematic uncertainties, $\mathcal{B}_{K^{*}\bar{K}}$ is determined to be $(0.1\pm1.2(\mathrm{stat})\pm1.1(\mathrm{syst}))\times 10^{-6}$, and $\mathcal{R}$ is determined to be $0.003\pm0.040(\mathrm{stat})\pm0.026(\mathrm{syst})$.
The $\mathcal{B}_{K^{*}\bar{K}}< 2.7\times 10^{-6}$ is obtained at the $90\%$ confidence level (C.L.) and $\mathcal{R}< 0.081$ at the $90\%$ C.L.
The suppression of the $X(2370)$ decay into $K^{*}(892)\bar{K}$ supports that the $X(2370)$ is a flavor-singlet. The $X(2370)$ is the first flavor-singlet light hadron observed above $1~\GeV/c^{2}$~\cite{pdg2024}, and it is consistent with the glueball interpretation.

In the $K_{S}^{0}K_{S}^{0}\pi^{0}$ and $\pi^{0}\pi^{0}\eta$ invariant mass spectra in $J/\psi\rightarrow\gamma K_{S}^{0}K_{S}^{0}\pi^{0}$ and $\gamma \pi^{0}\pi^{0}\eta$~\cite{combined_X2370}, 
the $X(2370)$ is the unique mass peak observed between $2.3~\GeV/c^{2}$ and $\eta_{c}$ mass peak, 
which indicates that other possible $0^{-+}$ mesons in this mass range are either with much lower production rates or with much more broad widths than the $X(2370)$. 
To obtain product BFs for $J/\psi\to\gamma X(2370)$ with $X(2370)\to K\bar{K}\pi$, $\pi\pi\eta$, and $\pi\pi\eta^{\prime}$, fits are performed to the $K_{S}^{0}K_{S}^{0}\pi^{0}$, $\pi^{0}\pi^{0}\eta$, and $\pi^{+}\pi^{-}\eta^{\prime}$ invariant mass spectra, with the mass and width of the $X(2370)$ fixed to the combined results~\cite{combined_X2370}. The systematic uncertainties from the detection efficiency, background model, PHSP parametrization, and additional resonance are considered. 
The resulting product BFs, together with the result for $J/\psi\to\gamma X(2370), X(2370)\to K\bar{K}\eta^{\prime}$ from Ref.~\cite{epjc.s10052_gongli}, are summarized in Table~\ref{tab:x2370_summary}.
As a reasonable estimate, the $\mathcal{B}[J/\psi\rightarrow\gamma X(2370)]$ should be greater than $1\times10^{-3}$, 
which means the $X(2370)$ has high production rate in $J/\psi$ radiative decays. 

\begin{table}[tb]
\centering
\renewcommand\arraystretch{1.2}
\setlength{\tabcolsep}{2.5mm}
\caption{Summary of measured product BFs for the $X(2370)$. The first uncertainties are statistical and the second are systematic. 
The results for $K\bar{K}\pi$, $\pi\pi\eta$, and $\pi\pi\eta^{\prime}$ are obtained from fits to the corresponding reconstructed invariant mass spectra, with isospin factors of 12, 3, and 1.5 applied, respectively.
The result for $K\bar{K}\eta^{\prime}$ is obtained by combining the $K_{S}^{0}K_{S}^{0}\eta^{\prime}$ (with isospin factor of 4) and $K^{+}K^{-}\eta^{\prime}$ (with isospin factor of 2) modes. 
}
% \resizebox{\columnwidth}{!}{

\begin{tabular}{lc}
\hline\hline
Decay channel &$\mathcal{B}$~$(\times10^{-4})$ \\
\hline

$J/\psi\to\gamma X(2370)\to \gamma K\bar{K}\pi$ & $3.25\pm0.25^{+0.73}_{-0.75}$ \\

$J/\psi\to\gamma X(2370)\to\gamma\pi\pi\eta$ & $3.2\pm0.1^{+0.9}_{-1.0}$ \\

$J/\psi\to\gamma X(2370)\to\gamma\pi\pi\eta^{\prime}$ & $1.94\pm0.04^{+0.33}_{-0.88}$ \\

$J/\psi\to\gamma X(2370)\to \gamma K\bar{K}\eta^{\prime}$ & $0.39\pm0.05\pm0.10$ \\

\hline\hline
\end{tabular}
% }
\label{tab:x2370_summary}
\end{table}

It is noticed that some phenomenological study calculated the production rate of $J/\psi\rightarrow\gamma X(2370)$ to be greater than $2.87\times10^{-3}$~\cite{Sun:2021kka}, 
but their calculation yielded that the decay BF of the $X(2370)$ to $K\bar{K}\pi$ mode is more than 50 times greater than that of the $K\bar{K}\eta^{\prime}$ mode, 
which is inconsistent with the above \mbox{BESIII} BF measurements of the $X(2370)$. 
Nevertheless, the high production rate of the $X(2370)$ in $J/\psi$ radiative decays is favored by the $0^{-+}$ glueball interpretation~\cite{Chen:2026lki,Zhang:2021xvl,Qin:2017qes}.

It should be noted that although the $X(2370)$ is a flavor-singlet and its decay modes should be similar to those of $\eta_{c}$, 
it is hard to directly compare the corresponding BF ratios of different decay modes between the $X(2370)$ and the $\eta_{c}$, since $\eta_{c}$ has much larger PHSP than the $X(2370)$ in all decay final states. 
For example, 
the 2D mass plot of $M_{K_{S}^{0}K_{S}^{0}}$ versus $M_{K_{S}^{0}K_{S}^{0}\eta^{\prime}}$ in the analysis of $J/\psi\rightarrow\gamma K_{S}^{0}K_{S}^{0}\eta^{\prime}$~\cite{2370jpc} shows that $\eta_{c}$ may dominantly decay into $f_{0}(1710)\eta^{\prime}\rightarrow K_{S}^{0}K_{S}^{0}\eta^{\prime}$, 
but there is no PHSP for the $X(2370)$ decaying to $f_{0}(1710)\eta^{\prime}$. 
Another example is that some multi-pion modes such as $\eta_{c}\rightarrow f_{2}(1270)f_{2}(1270)$ are allowed for the $\eta_{c}$ decays~\cite{pdg2024}, but they may be suppressed for the $X(2370)$ decays also due to PHSP limitations.

Given the upper limit on $\mathcal{B}[J/\psi\rightarrow\gamma X(2370)] \times \mathcal{B}[ X(2370) \rightarrow K^{*}(892)^{0}\bar{K}^{0}+\mathrm{c.c.} \rightarrow K^{0}_{S}K^{0}_{S}\pi^{0}] < 2.7\times 10^{-6}$ at the $90\%$ C.L., and $\mathcal{B}[J/\psi\rightarrow \gamma X(2370)]$ should be larger than $1\times10^{-3}$, we have $\mathcal{B}[X(2370)\rightarrow K^{*}(892)\bar{K}]<1.6\%$. Then, the partial width of the $X(2370)$ to $K^{*}(892)\bar{K}$ should be smaller than $2~\MeV$.
The suppression of the $K^{*}(892)\bar{K}$ decay mode indicates that the $X(2370)$ is inconsistent with the $\eta$-$\eta^{\prime}$ excitation interpretation, for which the partial width to $K^{*}(892)\bar{K}$ is expected to be $15\text{--}200~\MeV$~\cite{Yu:2011ta,Wang:2017iai}. This expectation is also consistent with a na\"ive extrapolation from the $\eta(1405)/\eta(1475)$ partial width to $K^{*}(892)\bar{K}$, by comparing the PHSP contributions for $\eta(1405)/\eta(1475)\to K^{*}(892)\bar{K}$ and $X(2370)\to K^{*}(892)\bar{K}$. Since $K^{*}(892)\bar{K}$ is one of the major decay modes of the $\eta(1405)/\eta(1475)$, the corresponding partial width is expected to be larger than $10~\MeV$. Moreover, the $P$-wave PHSP for an $\eta$-$\eta^{\prime}$ excitation around $2.37~\GeV/c^{2}$ to $K^{*}(892)\bar{K}$ is more than 30 times larger than that for the $\eta(1405)/\eta(1475)$. Therefore, from this simple extrapolation, the partial width of an $\eta$-$\eta^{\prime}$ excitation around $2.37~\GeV/c^{2}$ decaying to $K^{*}(892)\bar{K}$ would also be expected to be much larger than $20~\MeV$.

In addition, the flavor-singlet property of the $X(2370)$ disfavors $q\bar{q}$ interpretation, 
because no flavor-singlet state formed via $u\bar{u}+d\bar{d}$ and $s\bar{s}$ mixing has been predicted around $2~\GeV/c^{2}$ by the LQCD calculations~\cite{Dudek:2011tt}.  
From experimental data~\cite{pdg2024}, except for the $X(2370)$, no other flavor-singlet light meson has been observed above $1~\GeV/c^{2}$, 
which also supports the LQCD calculation. 

The $q\bar{q}$ interpretation is also disfavored by the strong suppressions of the $X(2370)$ decays into $\gamma\omega$ or $\gamma\phi$. 
Compared with radiative decays of other $0^{-+}$ mesons, such as the $\eta(1405)$ and $X(1835)$, as shown in Table~\ref{tab:prd111_resonances},  
the strong suppressions of $X(2370)\rightarrow\gamma\phi$, $\gamma\omega$, 
indicate that both $s\bar{s}$ and $u\bar{u}+d\bar{d}$ contents inside the 
$X(2370)$ should be very small. 
The upper limits on the $X(2370)$ in \mbox{Table~\ref{tab:prd111_resonances}} are consistent with the theoretical expectations of the pseudoscalar glueball~\cite{Hechenberger:2023ljn}. 

\begin{table}[tb]
\centering
\renewcommand\arraystretch{1.2}
\setlength{\tabcolsep}{2.5mm}
\caption{BFs or their upper limits at 90\% C.L. of the intermediate resonances measured in
$J/\psi \to \gamma\gamma\omega$ and $J/\psi \to \gamma\gamma\phi$. The first uncertainties are statistical and the second are systematic. }
\begin{tabular}{ccc}
\hline\hline
Resonance $R$
& \makecell{$J/\psi \to \gamma R \to \gamma\gamma\omega$\\$(\times10^{-6})$~\cite{BESIII:ggomega}}
& \makecell{$J/\psi \to \gamma R \to \gamma\gamma\phi$\\$(\times10^{-6})$~\cite{BESIII:2024ein}} \\
\hline

$\eta(1405)$
& $3.54 \pm 0.20^{+0.65}_{-0.24}$
& $3.57 \pm 0.18^{+0.59}_{-0.61}$ \\

$X(1835)$
& $< 0.11$
& $3.37 \pm 0.19^{+0.78}_{-1.10}$ \\

$X(2370)$
& $< 0.04$
& $< 0.11$ \\
\hline\hline

\end{tabular}
\label{tab:prd111_resonances}
\end{table}

Furthermore, the flavor-singlet property of the $X(2370)$ means that the $X(2370)$ should decay flavor symmetrically to many $S$-wave quasi-two-body modes in $\pi\pi\eta^{\prime}$, 
$K\bar{K}\eta^{\prime}$, $\pi\pi\eta$ and $K\bar{K}\pi$ processes, 
such as $f_{0}(980) \eta^{\prime}$, $a_{0}(980)\pi^{0}$, $K^{*}_{0}(1430)\bar{K}$, $f_{0}(1500) \eta$ (as indicated in the 2D mass plots in $J/\psi\rightarrow\gamma K_{S}^{0}K_{S}^{0}\eta^{\prime}$, 
$J/\psi\rightarrow\gamma K_{S}^{0}K_{S}^{0}\pi^{0}$ and $J/\psi\rightarrow\gamma\pi^{0}\pi^{0}\eta$~\cite{combined_X2370}). 
Therefore, the BF of each quasi-two-body modes is expected to be only of order $1\%\text{--}10\%$, which is also indicated by the similarities to $\eta_{c}$ decays, i.e., 
their partial widths are of the order of a few~$\MeV$ or even smaller, which is much smaller than typical OZI allowed decay partial width ($\sim100~\MeV$).
Such small partial widths strongly disfavor the interpretation of normal $q\bar{q}$ states, multi-quark states or hybrid decays~\cite{Sun:2021kka,Yu:2011ta,Dong:2020okt,Wang:2025nme,Su:2022eun}, since all these hadrons have light quark content in the initial states, so that their decays are expected to be dominated by OZI allowed modes~\cite{Wang:2017iai}.
In contrast, the narrow partial widths are expected by conventional understanding of glueballs, since their decays to $q\bar{q}$ states are OZI suppressed and the suppression only acts at one vertex because of the absence of initial $q\bar{q}$ annihilation for a glueball decay~\cite{Robson:1977pm}.
% This suppression is also predicted by the $1/N_{c}$ expansion, where $N_{c}$ is the number of colors, because to lowest order in $1/N_{c}$, glueballs are decoupled from $q\bar{q}$ states~\cite{Witten:1979kh}.
% Carlson:1980kh
For example, from the so called $\sqrt{\text{OZI}}$ rule~\cite{Robson:1977pm}, it can be na\"ively estimated that the partial decay width $\Gamma(G_{0^{-+}}\rightarrow\pi\pi\eta)$ is about $\sqrt{\Gamma(\text{OZI allowed})\times\Gamma(\eta_{c}\rightarrow\pi\pi\eta)}\sim\sqrt{(100-200)~\MeV\times0.5~\MeV}$, i.e., of the order of a few~${\rm MeV}$. Therefore, the narrow partial widths of the $X(2370)$ are also consistent with the glueball interpretation.

It is noticed that the $X(2370)$ mass is close to the $\Sigma\bar{\Sigma}$ mass threshold 
and it might be conjectured as a $\Sigma\bar{\Sigma}$ baryonium state. 
However, the production rate of the $X(2370)$ disfavors this interpretation given that the BFs of $J/\psi$ radiative decays to baryon pairs, such as $p\bar{p}$ and $\Lambda\bar{\Lambda}$, are only of the order $10^{-4}$ or even smaller~\cite{pdg2024}. Also there is no distortion observed at the $\Sigma\bar{\Sigma}$ mass threshold on the $X(2370)$ line shape, which indicates that the $X(2370)$ has no strong coupling with $\Sigma\bar{\Sigma}$. In addition, such a baryonium state is not a flavor-singlet as it should be for the $X(2370)$.

The upper limits of strong suppressions of $X(2370)\to K^{*}K, \gamma\omega, \gamma\phi$ may help to constrain the $q\bar{q}$ contributions inside the $X(2370)$. However, to obtain the precise fractions of the $q\bar{q}$ components, more studies are needed on the decay BFs of the $X(2370)$, including partial wave analyses on the possible interferences between the $X(2370)$ and nearby resonances such as $X(2120)$, $X(2260)$ and $X(2600)$ in various final states with considerations on possible threshold effects in this mass region. In addition, searching for $\omega\omega$, $\phi\phi$, $\omega\phi$ and $K^{*}(1410)\bar{K}$ modes may provide further confirmations on the flavor-singlet property of the $X(2370)$, since for a flavor-singlet state, it is expected to decay flavor symmetrically to $\omega\omega$ and $\phi\phi$, while $\omega\phi$ mode should be suppressed by OZI rule and $K^{*}(1410)\bar{K}$ mode is also forbidden by generalized $G$-parity conservation.

With 10 billion $J/\psi$ events collected at \mbox{BESIII}, the $X(2370)$ properties have been systematically studied. 
A dominant $0^{-+}$ glueball component is essential for a natural and complete explanation of the properties of the $X(2370)$~\textemdash~the mass, spin-parity, high production rate in $J/\psi$ radiative decays, decay property similarities to those of $\eta_{c}$, flavor-singlet property, narrow partial decay width, and suppression of radiative decays to $\omega$ and $\phi$, since these properties are all consistent with the features of a $0^{-+}$ glueball while other interpretations at present can hardly explain them simultaneously, especially on the flavor-singlet property, narrow partial decay widths and suppressions of radiative decays to $\omega$ and $\phi$. This supports that the lightest $0^{-+}$ glueball is the dominant constituent of the $X(2370)$.

We would like to thank Profs. K.T. Chao, Y. Chen, F.-K. Guo, T. Huang, X.Liu, J.-P. Ma, C.-F. Qiao, J.-J. Wu, Y.-B. Yang, Q. Zhao for their helpful discussions.

%% Saved at => 2026-06-13

The \mbox{BESIII} Collaboration thanks the staff of BEPCII (https://cstr.cn/31109.02.BEPC) and the IHEP computing center for their strong support. This work is supported in part by National Key R\&D Program of China under Contracts Nos. 2025YFA1613900, 2023YFA1606000, 2023YFA1606704; National Natural Science Foundation of China (NSFC) under Contracts Nos. 11635010, 11935015, 11935016, 11935018, 12025502, 12035009, 12035013, 12061131003, 12192260, 12192261, 12192262, 12192263, 12192264, 12192265, 12221005, 12225509, 12235017, 12342502, 12361141819, 12535005; the Chinese Academy of Sciences (CAS) Large-Scale Scientific Facility Program; the Strategic Priority Research Program of Chinese Academy of Sciences under Contract No. XDA0480600; CAS under Contract No. YSBR-101; 100 Talents Program of CAS; The Institute of Nuclear and Particle Physics (INPAC) and Shanghai Key Laboratory for Particle Physics and Cosmology; Agencia Nacional de Investigaci\'on y Desarrollo de Chile (ANID), Chile under Contract No. ANID CCTVal CIA250027; Istituto Nazionale di Fisica Nucleare, Italy; Knut and Alice Wallenberg Foundation under Contracts Nos. 2021.0174, 2021.0299, 2023.0315; Ministry of Development of Turkey under Contract No. DPT2006K-120470; National Research Foundation of Korea under Contract No. RS-2026-25486791; National Science and Technology fund of Mongolia; Polish National Science Centre under Contract No. 2024/53/B/ST2/00975; STFC (United Kingdom); Swedish Research Council under Contract No. 2019.04595; U. S. Department of Energy under Contract No. DE-FG02-05ER41374.

\bibliographystyle{apsrev4-2}
\bibliography{draft}% Produces the 

\onecolumngrid

\begin{small}
\begin{center}

% Separator between references and author list
\vspace*{0.35cm}
\rule{0.24\textwidth}{0.9pt}
\par\vspace{0.45cm}

M.~Ablikim$^{1}$\BESIIIorcid{0000-0002-3935-619X},
M.~N.~Achasov$^{4,c}$\BESIIIorcid{0000-0002-9400-8622},
P.~Adlarson$^{85}$\BESIIIorcid{0000-0001-6280-3851},
X.~C.~Ai$^{91}$\BESIIIorcid{0000-0003-3856-2415},
C.~S.~Akondi$^{32A,32B}$\BESIIIorcid{0000-0001-6303-5217},
R.~Aliberti$^{40}$\BESIIIorcid{0000-0003-3500-4012},
A.~Amoroso$^{84A,84C}$\BESIIIorcid{0000-0002-3095-8610},
Q.~An$^{80,67,\dagger}$,
M.~S.~Anderson$^{40}$\BESIIIorcid{0009-0008-1550-2632},
Y.~Bai$^{65}$\BESIIIorcid{0000-0001-6593-5665}
% ...
\\
\vspace{0.2cm}
(BESIII Collaboration)\\
\vspace{0.2cm} {\it
$^{1}$ Institute of High Energy Physics, Beijing 100049, People's Republic of China\\
% ...
$^{4}$ Budker Institute of Nuclear Physics SB RAS (BINP), Novosibirsk 630090, Russia\\
% ...
$^{24}$ Henan University of Technology, Zhengzhou 450001, People's Republic of China\\
% ...
$^{32}$ INFN Sezione di Ferrara, (A)INFN Sezione di Ferrara, I-44122, Ferrara, Italy; (B)University of Ferrara, I-44122, Ferrara, Italy\\
% ...
$^{40}$ Johannes Gutenberg University of Mainz, Johann-Joachim-Becher-Weg 45, D-55099 Mainz, Germany\\
% ...
$^{65}$ Southeast University, Nanjing 211100, People's Republic of China\\
% ...
$^{67}$ State Key Laboratory of Particle Detection and Electronics, Beijing 100049, Hefei 230026, People's Republic of China\\
% ...
$^{80}$ University of Science and Technology of China, Hefei 230026, People's Republic of China\\
% ...
$^{84}$ University of Turin and INFN, (A)University of Turin, I-10125, Turin, Italy; (B)University of Eastern Piedmont, I-15121, Alessandria, Italy; (C)INFN, I-10125, Turin, Italy\\
$^{85}$ Uppsala University, Box 516, SE-75120 Uppsala, Sweden\\
% ...
$^{91}$ Zhengzhou University, Zhengzhou 450001, People's Republic of China\\

\vspace{0.2cm}
$^{\dagger}$ Deceased\\
% ...
$^{c}$ Also at the Novosibirsk State University, Novosibirsk, 630090, Russia\\
% ...

}
%% ends here %%

\end{center}
\vspace{0.4cm}
\end{small}
% }

\twocolumngrid

\end{document}